\documentclass{ws-p8-50x6-00}

\begin{document}

\title{Multipole Analysis of a Benchmark Data Set for Pion Photoproduction}

\author{ R.A. Arndt$^1$, I. Aznauryan$^2$, R.M. Davidson$^3$, D.~Drechsel$^4$,
O.~Hanstein$^4$, S.S.~Kamalov$^5$, A.S.~Omelaenko$^6$, I.~Strakovsky$^1$,
L.~Tiator$^4$, R.L. Workman$^1$, S.N.~Yang$^7$}
\address{$^1$ Department of Physics,
The George Washington University, Washington, D.C. 20052, U.S.A\\
$^2$ Yerevan Physics Institute, Alikhanian Brothers St. 2, Yerevan, 375036 Armenia\\
$^3$ Department of Physics, Applied Physics and Astronomy, Rensselaer Polytechnic
Institute, Troy, NY 12180, U.S.A\\
$^4$ Institut f\"ur Kernphysik, Universit\"at Mainz, 55099 Mainz, Germany\\
$^5$ Laboratory of Theoretical Physics, JINR, 141980
Dubna, Russia\\
$^6$ National Science Center, Kharkov Institute of Physics and Technology,
Akademicheskaya St., 1, Kharkov, 61108, Ukraine\\
$^7$Department of Physics, National
Taiwan University, Taipei, Taiwan}

\maketitle

\abstracts{ We have fitted low- and medium-energy benchmark
datasets employing methods used in the MAID/SAID and dynamical
model analyses. Independent fits from the Mainz,
RPI, Yerevan, and Kharkov groups
have also been performed over the low-energy region. Results
for the multipole amplitudes are compared in order to gauge
the model-dependence of such fits, given identical data and
a single method for error handling.}

\section{Overview}

The following report summarizes results from a series of
fits to selected datasets for pion photoproduction,
a project initiated by the Partial-Wave Analysis Working Group
of BRAG (the Baryon Resonance Analysis Group). The goal
of this work was an evaluation of the model-dependence inherent
in multipole analyses of photoproduction data. In the past,
groups have constructed their own databases, and the
resulting differences have been
shown to significantly effect some multipoles, in particular
the E2/M1 ratio for the $\Delta (1232)$. The handling of
systematic errors has also differed, the common choices being
to either ignore them entirely, to combine them in quadrature with
statistical errors, or to use them in `floating' the angular
distributions.

The construction of low-energy (180-450 MeV) and medium-energy
(180-1200 MeV) datasets was carried out mainly for practical reasons.
Many groups have studied the region below 450 MeV, which is
dominated by the $\Delta (1232)$. Only a few groups have extended
their analyses over the full resonance region. As the recent
MAID and dynamical model fits (see the first contribution)
extend to 1 GeV, an upper limit of 1.2 GeV was chosen to
include as many independent fits as possible.

The fitted data are listed on the BRAG website\cite{BRAG} and
include differential cross section, target asymmetry ($T$), and
photon asymmetry $(\Sigma)$ data from proton targets only.
The low- and medium-energy datasets were
constructed to contain mainly recent measurements, with the goal
of minimizing redundancies and simplifying the fitting procedure.
In order to further simplify the exercise, systematic errors
were not included in the fits. This should be kept in mind when
$\chi^2$ values are quoted.

Below we have compiled the reports of individual groups,
giving details of the methods used, comments on the fit quality,
and ways these results could be interpreted. Finally, in
the last section, we summarize the findings of this study
and suggest ways it could be improved or extended.

\section{Multipole Analysis with MAID and a Dynamical Model}
\centerline{[ S.S. Kamalov, D. Drechsel, L. Tiator, and S.N. Yang ]}
\vskip .5cm

During the last few years we have developed and extended two
models for the analysis of pion photo and electroproduction, the
Dynamical Model~\cite{KY99} (hereafter called Dubna-Mainz-Taipei
(DMT) model) and the Unitary Isobar Model~\cite{MAID98} (hereafter
called MAID). The final aim of such an analysis is to shed more
light on the dynamics involved in nucleon resonance excitations
and to extract $N^*$ resonance properties in an unambiguous way.
For this purpose as a testing ground we will use benchmark data
bases recently created and distributed among different theoretical
groups.

The crucial point in a study of $N^*$ resonance properties is the
separation of the total amplitude (in partial channel
$\alpha=\{l,j\}$)
\begin{eqnarray}
t_{\gamma\pi}^{\alpha}=t_{\gamma\pi}^{B,\alpha}+t_{\gamma\pi}^{R,\alpha}
\end{eqnarray}
in background $t_{\gamma\pi}^{B,\alpha}$ and resonance
$t_{\gamma\pi}^{R,\alpha}$ contributions. In different theoretical
approaches this procedure is different, and consequently this
could lead to different treatment of the dynamics of $N^*$
resonance excitation. As an example we will consider the two
different models: DMT and MAID.

In accordance with Ref.\cite{KY99}, in the DMT model the
$t_{\gamma\pi}^{B,\alpha}$ amplitude is defined as
\begin{eqnarray}
t^{B,\alpha}_{\gamma\pi}(DMT)=e^{i\delta_{\alpha}}\cos{\delta_{\alpha}}
\left[v^{B,\alpha}_{\gamma\pi} + P\int_0^{\infty} dq'
\frac{q'^2\,R_{\pi
N}^{(\alpha)}(q_E,q')\,v^{B,\alpha}_{\gamma\pi}(q')}{E-E_{\pi
N}(q')}\right]\,,
\label{eq:backgr}
\end{eqnarray}
where $\delta_{\alpha}(q_E)$ and $R_{\pi N}^{(\alpha)}$ are the
$\pi N$ scattering phase shift and full $\pi N$ scattering
reaction matrix, in channel $\alpha$, respectively, $q_E$ is the
pion on-shell momentum. The pion photoproduction potential
$v^{B,\alpha}_{\gamma\pi}$ is constructed in the same way as in
Ref.\cite{MAID98} and contains contributions from the Born terms
with an energy dependent mixing of pseudovector-pseudoscalar
(PV-PS) $\pi NN$ coupling and t-channel vector meson exchanges. In
the DMT model $v^{B,\alpha}_{\gamma\pi}$ depends on 7 parameters:
The PV-PS mixing parameter $\Lambda_m$ (see Eq.(12) of
Ref.\cite{MAID98}), 4 coupling constants  and 2 cut-off parameters
for the vector mesons exchange contributions.

In the extended version of MAID, the $s$, $p$, $d$ and $f$ waves
of the background amplitudes $t_{\gamma\pi}^{B,\alpha}$ are
defined in accordance with the K-matrix approximation
\begin{equation}
 t_{\gamma\pi}^{B,\alpha}({\rm MAID})=
 \exp{(i\delta_{\alpha})}\,\cos{\delta_{\alpha}}
 v_{\gamma\pi}^{B,\alpha}(q,W,Q^2)\,,
\label{eq:bg00}
\end{equation}
where $W\equiv E$ is the total $\pi N$ c.m. energy and
$Q^2=-k^2>0$ is the square of the virtual photon 4-momentum. Note
that in actual calculations, in order to take account of inelastic
effects, the factor $\exp{(i\delta_{\alpha})}
\cos{\delta_{\alpha}}$ in Eqs.(2-3) is replaced by
$\frac{1}{2}[\eta_{\alpha}\exp{(2i\delta_{\alpha})} +1]$, where
both the $\pi N$ phase shifts $\delta_{\alpha}$ and inelasticity
parameters $\eta_{\alpha}$ are taken from the analysis of the SAID
group\cite{VPI97}.

From Eqs. (\ref{eq:backgr}) and (\ref{eq:bg00}), one finds that
the difference between the background terms of MAID and of the DMT
model is that pion off-shell rescattering contributions (principal
value integral) are not included in the background of MAID. From
our previous studies of the $p$-wave multipoles in the (3,3)
channel~\cite{KY99} it follows that they are effectively included
in the resonance sector leading to the dressing of the $\gamma
N\Delta$ vertex. However, in the case of $s$ waves the DMT results
show that off-shell rescattering contributions are very important
for the $E_{0+}$ multipole in the $\pi^0p$ channel. In this case
they have to be taken into account explicitly. Therefore, in the
extended version of MAID we have introduced a new phenomenological
term in order to improve the description of the $\pi^0$
photoproduction at low energies,
\begin{eqnarray}
E_{corr}(MAID)= \frac{\Delta E}{(1 + B^2q^2_{E})^2}\,F_D(Q^2)\,,
\label{eq:corr}
\end{eqnarray}
where $F_D$ is the standard nucleon dipole form factor, $B=0.71$
fm and $\Delta E$ is a free parameter which can be fixed by
fitting the low energy $\pi^0$ photoproduction data. Thus the
background contribution in MAID finally depends on 8 parameters.
Below $\pi^+n$ threshold for both models we also take into account
the cusp effect due to unitarity, as it was described in
Ref.\cite{Larget}, i.e.
\begin{eqnarray} E_{cusp}= - a_{\pi N} \,\omega_c\,{\rm Re}
E_{0+}^{\gamma\pi^+}\,\sqrt{1-\frac{\omega^2}{\omega_c^2}}\,,
\label{eq:cusp}
\end{eqnarray}
where $\omega$ and $\omega_c=140$ MeV are the $\pi^+$ c.m.
energies corresponding to $W=E_p + E_{\gamma}$ and
$W_c=m_n+m_{\pi^+}$, respectively, and $a_{\pi N}=0.124/m_{\pi^+}$
is the pion charge exchange amplitude.

For the resonance contributions, following  Ref.\cite{MAID98}, in
both models the Breit-Wigner form is assumed, i.e.
\begin{equation}
t_{\gamma\pi}^{R,\alpha}(W,Q^2)\,=\,{\bar{\cal
A}}_{\alpha}^R(Q^2)\, \frac{f_{\gamma R}(W)\Gamma_R\,M_R\,f_{\pi
R}(W)}{M_R^2-W^2-iM_R\Gamma_R} \,e^{i\phi_R}\,,
\label{eq:BW}
\end{equation}
where $f_{\pi R}$ is the usual Breit-Wigner factor describing the
decay of a resonance $R$ with total width $\Gamma_{R}(W)$ and
physical mass $M_R$. The phase $\phi_R(W)$ in Eq. (\ref{eq:BW}) is
introduced to adjust the phase of the total multipole to equal the
corresponding $\pi N$ phase shift $\delta_{\alpha}$.

The main subject of our study in the resonance sector is the
determination of the strengths of the electromagnetic transitions
described by the amplitudes ${\bar{\cal A}}_{\alpha}^R(Q^2)$. In
general, they are considered as free parameters which have to be
extracted from the analysis of the experimental data. In our two
models we have included contributions from the 8 most important
resonances, listed in the Table 1. The total number of ${\bar{\cal
A}}_{\alpha}^R$ amplitudes is 12 and they can be expressed also in
terms of the 12 standard helicity elements $A_{1/2}$ and
$A_{3/2}$.
\begin{table}[htbp]
\begin{center}
\begin{tabular}{|cc|c|c|c|c|}
\hline $N^*$ &  &  MAID & MAID    &  DMT    & PDG2000   \\
      &  &  current    & HE fit & HE fit &  \\
\hline $P_{33}(1232)$ & $A_{1/2}$ & -138  & -143  & ---  &
-135$\pm$ 6 \\
               & $A_{3/2}$ & -256  & -264  & ---  & -255$\pm$ 8 \\
\hline $P_{11}(1440)$ & $A_{1/2}$ &  -71  & -81   &  -77  &
-65$\pm$ 4 \\ \hline $D_{13}(1520)$ & $A_{1/2}$ &  -17  & -6    &
-7   &  -24$\pm$ 9 \\
               & $A_{3/2}$ &  164  & 160  & 165  &  166$\pm$ 5 \\
\hline $S_{11}(1535)$ & $A_{1/2}$ &   67  &  81   &  102  &
90$\pm$ 30 \\ \hline $S_{31}(1620)$ & $A_{1/2}$ &   0  &  86    &
37   &  27$\pm$ 11 \\ \hline $S_{11}(1650)$ & $A_{1/2}$ &   39  &
32   &  34   & 53$\pm$ 16  \\ \hline $F_{15}(1680)$ & $A_{1/2}$ &
-10  &  5    &  10   & -15$\pm$ 6  \\
               & $A_{3/2}$ &  138  &  137  &  132  & 133 $\pm$ 12\\
\hline $D_{33}(1700)$ & $A_{1/2}$ &   86  & 119   &  107  &
104$\pm$ 15 \\
               & $A_{3/2}$ &   85  &  82   &  74   & 85$\pm$  22 \\
\hline
PV-PS mixing:   & $\Lambda_m$ &  450   &  406    & 302  &     \\
\hline
       &  $\Delta E$ &   2.01 & 1.73 & ---   &  \\
\hline
       &  $ \chi^2$/d.o.f. &   11.5 & 6.10 &  6.10  &     \\
\hline
\end{tabular}
\end{center}
\caption{Proton helicity amplitudes (in $10^{-3}\,GeV^{-1/2}$),
values of the PV-PS mixing parameter $\Lambda_m$ (in MeV) and
low-energy correction parameter $\Delta E$ (in
$10^{-3}/m_{\pi^+}$) obtained after the high-energy (HE) fit}
\end{table}
Thus, to analyze experimental data we have a total of 19
parameters in DMT and 20 parameters in MAID. The final results
obtained after the fitting of the high-energy (HE) benchmark data
base with 3270 data points in the photon energy range $180 <
E_{\gamma}<1200$ MeV are given in Table 1.

For the  analysis of the low-energy (LE) data base with 1287 data
points in the photon energy range  $180 < E_{\gamma}<450$ MeV in
the DMT model we used only 4 parameters: The PV-PS mixing
parameter and 3 parameters for the $P_{33}(1232)$ and
$P_{11}(1440)$ resonances. In MAID we have one more parameter due
to the low energy correction given by Eq. (\ref{eq:corr}). The
final results for the helicity elements and the E2/M1 ratio (REM)
are given in Table 2.
\begin{table}[htbp]
\begin{center}
\begin{tabular}{|cc|c|c|c|c|}
\hline
$N^*$ &  &  MAID & MAID    &  DMT    & PDG2000   \\
      &  &  current    & LE fit & LE fit &  \\
\hline
$P_{33}(1232)$ & $A_{1/2}$ & -138  & -142  & ---  & -135$\pm$ 6 \\
               & $A_{3/2}$ & -256  & -265  & ---  & -255$\pm$ 8 \\
\hline
$P_{11}(1440)$ & $A_{1/2}$ &  -71  & -81   &  -93  &  -65$\pm$ 4 \\
\hline
       &  REM(\%)   &   -2.2   &  -1.9    & -2.1  & -2.5$\pm$ 0.5\\
\hline
       &  $ \chi^2$/d.o.f. &   4.76 & 4.56 &  3.59  &  \\
\hline
\end{tabular}
\end{center}
\caption{Proton helicity elements (in $10^{-3}\,GeV^{-1/2}$) and
REM=E2/M1 ratio (in \%) obtained from the LE fit.}
\end{table}
In Table 3 we summarize our results and show the $\chi^2$ obtained
for different channels and different observables after fitting the
LE and HE data bases.
\begin{table}[htbp]
\begin{center}
\begin{tabular}{|c|ccc|ccc|}
\hline
     &       & LE     &         &        & HE   &     \\
Observables & N & MAID   &    DMT
&  N &  MAID   &    DMT    \\ \hline
$\frac{d\sigma}{d\Omega}(\gamma,\pi^+)$  &
 317 & 4.68    &   3.32  & 871 & 6.36  &   5.95  \\
$\frac{d\sigma}{d\Omega}(\gamma,\pi^0)$  &
 354 & 7.22    &   5.74  & 859 & 6.87  &   5.85  \\
\hline $\Sigma (\gamma,\pi^+)$  &
 245  & 2.79   &   2.57  &  546 & 4.57  & 6.49  \\
$\Sigma (\gamma,\pi^0)$  &
 192  & 2.22   &   1.58  & 488 &  7.65  & 7.65  \\
\hline $T (\gamma,\pi^+)$&
 107  & 3.28    &   2.94  & 265 & 3.75 &  4.17  \\
$T (\gamma,\pi^0)$  &
72 & 5.18    &   4.84  &  241 & 5.31  & 5.65  \\
\hline
  Total  & 1287 & 4.56    &   3.64  & 3270 & 6.10  &   6.10  \\
\hline
\end{tabular}
\end{center}
\caption{$\chi^2/N$ for the cross sections
($\frac{d\sigma}{d\Omega}$), photon ($\Sigma$) and target ($T$)
asymmetries in $(\gamma,\pi^+)$ and $(\gamma,\pi^0)$ channels
obtained after LE and HE fit. $N$ is the number of data points}
\end{table}
Note that the largest $\chi^2$ in the LE fit we get for
differential cross sections and target asymmetries in
$p(\gamma,\pi^0)p$. Similar results were obtained practically in
all other analyses. A detailed comparison with the results of
different theoretical groups is given on the website
$http://gwdac.phys.gwu.edu/analysis/pr\_benchmark.html$. Below, in
Fig. 1 we show only one interesting example, the $E_{0+}$
multipole in the channel with total isospin 1/2. In this channel
contributions from the $S_{11}(1535)$ and $S_{11}(1620)$
resonances are very important. At $E_{\gamma}>750$ MeV our values
for the real part of the $_pE_{0+}^{1/2}$ amplitude are mostly
negative and lower than the results of the SAID multipole
analysis. The only possibility to remove such a discrepancy in our
two models would be to introduce a third $S_{11}$ resonance.
Another interesting result is related to the imaginary part of the
$_pE_{0+}^{1/2}$ amplitude and, consequently, to the value of the
helicity elements given in Table 1. Within the DMT model for the
$S_{11}(1535)$ we obtain $A_{1/2}=102$ for a total width of 120
MeV, which is more consistent with the results obtained in $\eta$
photoproduction, than with previous pion photoproduction results
obtained by the SAID and MAID groups.
\begin{figure}[tbp]
\begin{center}
\epsfig{file=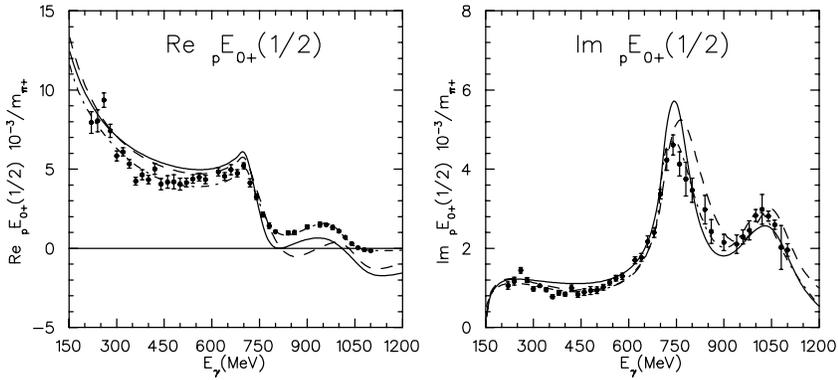, angle=90, width= 11 cm}
\end{center}
\caption{$_pE_{0+}^{1/2}$ multipole obtained after the HE fit
using MAID (solid curves) and DMT (dashed curves). The dash-dotted
curves and data points are the results of the global and
single-energy fits obtained by the SAID group.} \vspace*{-0.5cm}
%\caption{$_pE_{0+}^{1/2}$ multipole obtained after the HE fit
%using MAID (red curves) and DMT (blue curves). Green curves and
%data points are the results of global and single-energy fits
%obtained by the SAID group.} \vspace*{-0.5cm}
\end{figure}

\def\CPbar{\hbox{{\rm CP}\hskip-1.80em{/}}}%temp replacement due to no font
% Some useful journal names
\def\NCA{\em Nuovo Cimento}
\def\NIM{\em Nucl. Instrum. Methods}
\def\NIMA{{\em Nucl. Instrum. Methods} A}
\def\NPA{{\em Nucl. Phys.} A}
\def\NPB{{\em Nucl. Phys.} B}
\def\PLB{{\em Phys. Lett.}  B}
\def\PRL{\em Phys. Rev. Lett.}
\def\PR{\em Phys. Rev.}
\def\PRC{{\em Phys. Rev.} C}
\def\PRD{{\em Phys. Rev.} D}
\def\PHR{\em Phys. Rep.}
\def\ZP{\em Z. Phys.}
\def\ZPC{{\em Z. Phys.} C}

\section{Multipole Analysis of Pion Photoproduction with Constraints
  from Fixed-$t$ Dispersion Relations and Unitarity}

\centerline{ [ O. Hanstein, D. Drechsel and L. Tiator ]}
\vskip .5cm

\subsection{Outline of the Analysis}

The method presented in Ref.~\cite{ha98} has been used to analyze the
benchmark data set.

The starting point of our analysis is the fixed-$t$ dispersion relation
for the invariant amplitudes\cite{ba61},
\begin{equation}
  \label{fixed_t_photo}
  {\rm {Re}}A_{k}^{I}(s, t) = A_{k}^{I,\mathrm{pole}}(s, t)+
  \frac{1}{\pi}{\mathcal P}\!\!\int\limits_{s_{\mathrm{thr}}}^{\infty}
  ds'\left(\frac{1}{s'-s}+\frac{\varepsilon^{I}\xi_{k}}{s'-u}\right)
  {\rm {Im}}A_{k}^{I}(s', t),
\end{equation}
with the Mandelstam variables $s$, $u$ and $t$, the isospin index
$I=0,\pm$, $s_{\mathrm{thr}} = (m_{N}+m_{\pi})^{2}$,
$\varepsilon^{I} = \pm 1$ and $\xi_{k} = \pm 1$. The pole terms
$A_{k}^{I,\mathrm{pole}}(s, t)$ are obtained by evaluating the Born
approximation in pseudoscalar coupling.

The multipole projection of the dispersion relations
(\ref{fixed_t_photo}) leads to a system of coupled integral equations of
the form
\begin{equation}
  \label{multi_int}
  {\rm {Re}}{\mathbf{\mathcal M}}_{l}^{I}(W) =
   {\mathbf{\mathcal M}}_{l}^{I,\mathrm{pole}}(W) + \frac{1}{\pi}{\mathcal P}
  \!\!\int\limits_{W_{\mathrm{thr}}}^{\infty}dW'\sum_{l'=0}^{\infty}
  {\mathbf{\mathcal K}}_{ll'}^{I}(W,W')
  {\rm {Im}}{\mathbf{\mathcal M}}_{l}^{I}(W'),
\end{equation}
where ${\mathcal M}_{l}^{I}(W)$ denotes any of the multipoles
$E_{l\pm}^{I}$ or $M_{l\pm}^{I}$. The integral kernels
${\mathcal K}_{ll'}^{I}(W,W')$ are regular
kinematical functions except for the diagonal kernels ${\mathcal K}_{ll}^{I}$,
which contain a term $\propto 1/(W-W')$.

\begin{figure}[htb]
\psfig{figure=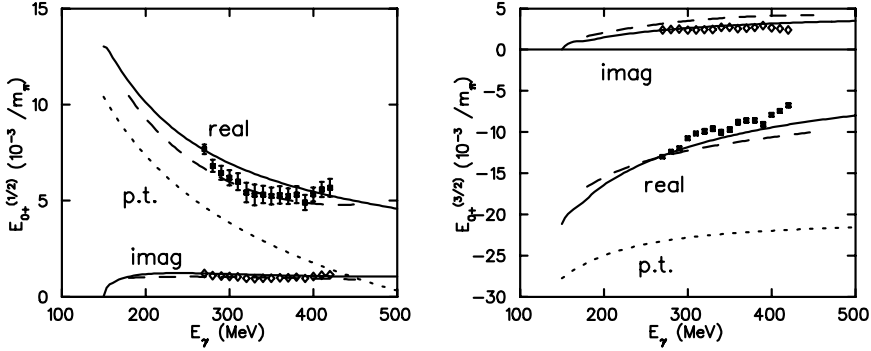,height=1.8in}
\caption{The real and imaginary parts and the pole term (p.t.) contributions
  of the amplitudes $E_{0+}(1/2)$ and $E_{0+}(3/2)$. The results of
  our fit (solid lines) are compared with those from
  Ref.~\protect\cite{Az01} (dashed lines). The data points are the
  result of our energy independent fit.
\label{fig:e0p}}
\end{figure}

\begin{figure}[htb]
\psfig{figure=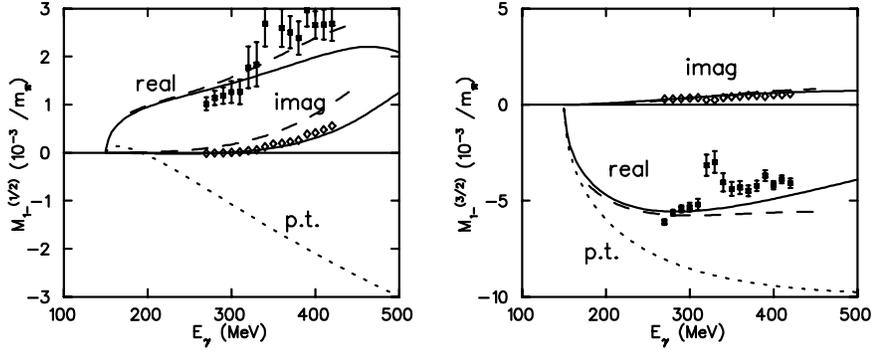,height=1.8in}
\caption{The real and imaginary parts and the pole term (p.t.) contributions
  of the amplitudes $M_{1-}(1/2)$ and $M_{1-}(3/2)$. Symbols as in
  Fig.~\protect\ref{fig:e0p}.
\label{fig:m1m}}
\end{figure}

\begin{figure}[htb]
\psfig{figure=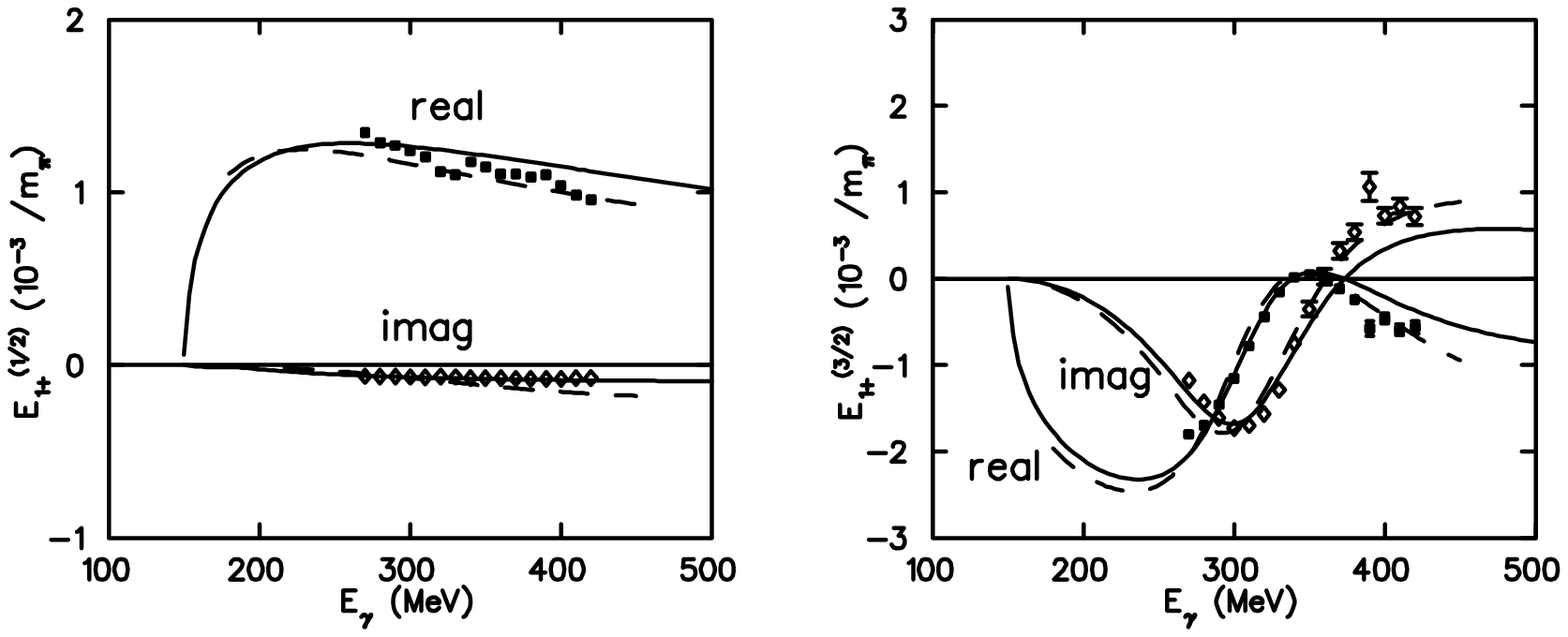,height=1.8in}
\caption{The real and imaginary parts
  of the amplitudes $E_{1+}(1/2)$ and $E_{1+}(3/2)$. Symbols as in
  Fig.~\protect\ref{fig:e0p}.
\label{fig:e1p}}
\end{figure}

\begin{figure}[htb]
\psfig{figure=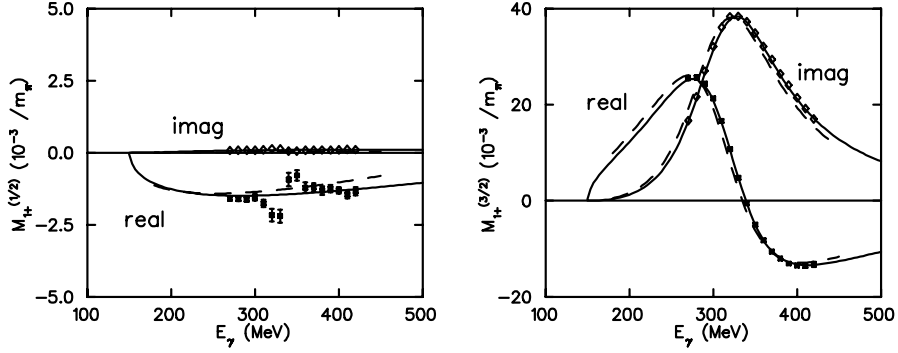,height=1.8in}
\caption{The real and imaginary parts
  of the amplitudes $M_{1+}(1/2)$ and $M_{1+}(3/2)$. Symbols as in
  Fig.~\protect\ref{fig:e0p}.
\label{fig:m1p}}
\end{figure}

After appropriate simplifications, namely neglect of weakly coupling
integral kernels and restriction to the partial waves of low angular
momentum, the integral equations (\ref{multi_int}) can be solved in
different ways. We based our analysis on the method of
Omn\`es\cite{om58}
because this method
leads to a parametrization of the multipoles in a natural way and
thus allows for analyzing experimental data. As an additional input,
the solution of the integral equations requires the phases
$\phi_{l}^{I}(W)$ of the
multipoles on the whole range of integration. According to the
Fermi-Watson theorem, these phases
are equal to the corresponding $\pi N$ scattering phase
shifts below 2$\pi$ threshold, $\phi_{l}^{I}(W) = \delta_{l}^{I}(W)$.
Above 2$\pi$ threshold, we use the ansatzes
\begin{equation}
  \label{uni_im_2}
  \phi_{l}^{I}(W)=\arctan\left(\frac{1-\eta_{l}^{I}(W)\cos 2\delta_{l}^{I}(W)}
  {\eta_{l}^{I}(W)\sin 2\delta_{l}^{I}(W)}\right).
\end{equation}
for the $S_{11}$, $P_{11}$, $P_{33}$, and $D_{13}$ waves, and
\begin{equation}
  \label{uni_im_1}
  \phi_{l}^{I}(W)=\arctan\left(\frac{\eta_{l}^{I}(W)\sin 2\delta_{l}^{I}(W)}
  {1+\eta_{l}^{I}(W)\cos 2\delta_{l}^{I}(W)}\right),
\end{equation}
for the $S_{31}$, $P_{13}$, and $P_{31}$ waves. These ansatzes
are each based on unitarity\cite{sc69} and an additional assumption.
They contain the scattering phases shifts $\delta_{l}^{I}(W)$ and the
inelasticity parameters $\eta_{l}^{I}(W)$ of $\pi N$ scattering.
Both ansatzes give $\phi_{l}^{I}(W) = \delta_{l}^{I}(W)$ below $2\pi$
threshold. Since partial wave analyses of $\pi N$ scattering are
only available up to about
$W = 2$ GeV, we cut the integrals off at this energy to avoid the
integration of unknown functions. Instead we represented the
contributions of the imaginary parts at higher energies
by $t$-channel exchange of vector
mesons. The integral equations then take the form
\begin{eqnarray}
\label{mul_int_fin}
{\mathrm{M}}_{l}^{I}(W) & = & {\mathrm{M}}_{l}^{I,\mathrm{pole}}(W) +
  \frac{1}{\pi}\int\limits_{W_{\mathrm{thr}}}^{\Lambda}\frac{h_{l}^{I\ast}(W')
  {\mathrm{M}}_{l}^{I}(W')dW'}{W'-W-i\varepsilon}
 \\ & &
 + \frac{1}{\pi}\sum_{l',I'}\int\limits_{W_{\mathrm{thr}}}
  ^{\Lambda}{\mathrm{K}}_{ll'}^{II'}(W,W')h_{l'}^{I'\ast}(W')
  {\mathrm{M}}_{l'}^{I'}(W')dW'
  + {\mathrm{M}}_{l}^{I,\mathrm{V}}(W), \nonumber
\end{eqnarray}
where  $\Lambda=2$ GeV.

The solutions relevant for our case are the sum of a particular
solution, which contains the inhomogeneities as driving terms, and a
solution to the homogeneous equation multiplied by an arbitrary real
coefficients $c_{l}^{I}$:
\begin{equation}
  \label{multi_sol}
  {\mathcal M}_{l}^{I}(W) = {\mathcal M}_{l}^{I,{\rm part}}(W)
                          + c_{l}^{I} {\mathcal M}_{l}^{I,{\rm hom}}(W).
\end{equation}
The coefficients $c_{l}^{I}$ are the fitting parameters in our
procedure. In addition, we varied the coupling constants of the
$\omega$ and the $\rho$ meson. Since the addition of the homogeneous
solution is only allowed for multipoles for which the phase is
different from zero in the asymptotic limit, we decided to fit the
parameters $c_{l}^{I}$ only for the following multipoles:
$E_{0+}(0)$, $E_{0+}(1/2)$, $M_{1-}(0)$, $M_{1-}(1/2)$, $E_{1+}(3/2)$
and $M_{1+}(3/2)$. So we end up with a 10 parameter fit, the results of
which are discussed in the next section.

\subsection{Results}
The fit to the benchmark data set leads to an overall $\chi^{2}$ of
3.7 per data point. This high value is mainly due to the differential
cross sections and target asymmetries of $\pi^{0}$ production (see
Table \ref{tab:chi2}).
\begin{table}[ht]
\caption{The $\chi^{2}$ per data point in our fit for the individual
observables. It is seen that, except for the beam asymmetry $\Sigma$,
the description of $\pi^{0}$ production data is much poorer than that
of $\pi^{+}$ production data.\label{tab:chi2}}
\vspace{0.4cm}
\begin{center}
\begin{tabular}{|l|cc|}
\hline
& $\pi^{0}p$  & $\pi^{+}n$ \\
\hline
$d\sigma/d\Omega$ & 7.4  & 2.8  \\
$\Sigma$ & 1.2  & 1.6  \\
$T$ & 5.2  & 2.9  \\
\hline
\end{tabular}
\end{center}
\end{table}

The magnitudes of the vector meson coupling constants as determined by
our fit differ somewhat from the values quoted in the literature. This can
be attributed to the fact that the dispersion integrals up to 2 GeV
already contain a certain fraction of the vector meson contributions
(see Table \ref{tab:vec}).

\begin{table}[ht]
\caption{Comparison between the vector meson coupling constants
  resulting from our fit and values quoted in the literature.\label{tab:vec}}
\vspace{0.4cm}
\begin{center}
    \begin{tabular}{|l|rrrr|}
    \hline
    & $g_{\rho}^{{\mathcal V}}$ & $g_{\rho}^{{\mathcal T}}$ &
      $g_{\omega}^{{\mathcal V}}$ & $g_{\omega}^{{\mathcal T}}$ \\
    \hline
    this work & 4.85 & 15.69 & 6.78 & -1.67 \\
    Ref.~\cite{Mac87} & 3.24 & 19.81 & 15.85 & 0 \\
    Ref.~\cite{Mer95} & 1.99 & 12.42 & 20.86 & -3.41 \\
    \hline
    \end{tabular}
\end{center}
\end{table}

Our results for the s- and p-wave multipoles are shown in
Figs.~\ref{fig:e0p} to \ref{fig:m1p}. As the procedure presented by Inna
Aznauryan\cite{Az01} at this workshop is closely related to ours, her results are
also shown for comparison. The real parts of the $E_{0+}$ and $M_{1-}$
multipoles differ significantly from the pole term contributions. In
our approach, this difference, which in other approaches has to be
provided by mechanisms like rescattering, is due to contributions from
dispersion integrals.

\section{Analysis of Low-Energy Benchmark Data Using Fixed-t Dispersion Relations}

\centerline{[ I.G.Aznauryan ]}
\vskip .5cm

In this analysis the low-energy benchmark data are
analyzed using fixed-t dispersion relations
within the approach which is close to the approach
developed in Refs. \cite{DeLi,DeLi2,DeLi3,AzSt}. The real parts
of the amplitudes are constructed through real parts of invariant
amplitudes $A_i^{(\pm,0)}(s,t),~i=1,4$ (Ref. \cite{Chew}).
which are obtained using fixed-t dispersion relations:
\begin{equation}
{\rm Re} A_i^{(\pm,0)}(s,t)=A_i^{Pole}(s,t)+
\frac{P}{\pi}\int\limits_{s_{thr}}^{s_{max}}
{\rm Im} A_i^{(\pm,0)}(s',t)\left(\frac{1}{s'-s}+\frac{\eta_i \eta^{(\pm,0)}}
{s'-u}\right)ds',
\label{1}\end{equation}
with
\begin{equation}
A_i^{Pole}(s,t) = A_i^{Born}(s,t)+A_i^{\omega}(s,t)+A_i^{\rho}(s,t)
\end{equation}
where $\eta^{(+,0)}=-\eta^{(-)}=1$, $\eta_1=\eta_2=-\eta_3=\eta_4=1$.
$A_i^{Born}(s,t)$ are the contributions of the nucleon poles
in the s- and u-channels and  of the pion pole in the t-channel.
In the dispersion relations\cite{DeLi}, the dispersion integrals
are taken over resonance energy region up to $s_{max}=(2~GeV)^2$,
and it is supposed that the integrals over higher
energies can be approximated by the t-channel $\omega$
and $\rho$-contributions: $A_i^{\omega}(s,t),~A_i^{\rho}(s,t)$.
These contributions are taken in the form presented in Ref. \cite{ha98}.
The integrals over resonance energy region are saturated by the resonances
used in the VPI analysis of pion photoproduction data  \cite{VPI97}. The coupling constants
for all resonances, except $P_{33}(1232)$, are taken from this analysis.
The resonance contributions are parametrized in the Breit-Wigner form according to
Ref.\cite{Walker}.

The multipole amplitudes $M_{1+}^{3/2}$ and $E_{1+}^{3/2}$
corresponding to the $P_{33}(1232)$ resonance are parametrized
using the approach developed in Refs. \cite{sc69,SRWK}.
According to this approach the amplitudes $M_{1+}^{3/2},~E_{1+}^{3/2}$
are the solutions of the singular integral equations which
follow from dispersion relations for these amplitudes, and
have the form:
\begin{equation}
M(s)=M_{part}^{Born}(s)+M_{part}^{\omega}(s)+
c_MM^{hom}(s),
\label{2}\end{equation}
where $M(s)$ denotes any of the amplitudes
$M_{1+}^{3/2},~E_{1+}^{3/2}$.
$M_{part}^{Born}(s)$ and $M_{part}^{\omega}(s)$ are the
particular solutions of the integral equations
generated by the Born and $\omega$ contributions.
Particular solutions have definite magnitudes fixed by
 these contributions.
 $M^{hom}(s)$ is the solution of the homogeneous part
 of the integral equation; it has a certain energy dependence
 fixed by the integral equation and an arbitrary weight
 which was found by fitting  the data.

 In the $P_{33}(1232)$ resonance region we have taken into account
 also the contributions of the non-resonant multipoles
 $E_{0+}^{1/2},~E_{0+}^{(0)},~E_{0+}^{3/2},~M_{1-}^{1/2},~M_{1-}^{(0)}$
 into ${\rm Im}A_i^{(\pm,0)}(s,t)$.
 These multipole amplitudes
 were found by calculating their real parts from the dispersion relations\cite{DeLi};
 then the imaginary parts of the multipole amplitudes
 at $W<1.3~GeV$ were found using the Watson theorem.
 At higher energies the smooth cutoff of these contributions was made.

\begin{table}
\begin{center}
\begin{tabular}{|c|c|}
\hline ${\rm Im} M_{1+}^{3/2}$&$52.458\pm 0.002~mFm$\\
\hline ${\rm Im} E_{1+}^{3/2}$&$-1.19\pm 0.01~mFm$\\
\hline $E2/M1$&-0.023\\
\hline
\end{tabular}
\end{center}
\caption{The amplitudes $M_{1+}^{3/2},E_{1+}^{3/2}$
at the resonance position.}
\end{table}

\vskip 10pt
\begin{table}
 \begin{center}
\begin{tabular}{|c|c|c|c|c|}
\hline
 &Present analysis&Ref.\cite{ha98}&Ref.\cite{MAID98}&Ref.\cite{Dumb}\\
\hline $g_{\omega}^V$&4&6.8&21&8-14\\
\hline $g_{\omega}^T$&-24&-1.7&-12&0-(-14)\\
\hline $g_{\varrho}^V$&2.5&4.9&2&1.8-3.2\\
\hline $g_{\varrho}^T$&21&16&13&8-21\\
\hline
\end{tabular}
\end{center}
\caption{The $\omega NN$ and $\varrho NN$ coupling constants
in comparison with results from other sources.}
\end{table}

\vskip 10pt
\begin{table}
\begin{center}
\begin{tabular}{|c|c|c|c|c|c|c|}
\hline
 &$\sigma(\pi^+)$&$A(\pi^+)$&$T(\pi^+)$&$\sigma(\pi^0)$&$A(\pi^0)$&$T(\pi^0)$\\
\hline $\chi 2/N$&922/317&440/253&245/107&2568/347&276/192&394/71\\
\hline
\end{tabular}
\end{center}
\caption{The values of $\chi 2$.}
\end{table}

Our fitting parameters were:

 (1) the constants $c_M$ and $c_E$
 which correspond to the magnitudes of homogeneous solutions
 for  $M_{1+}^{3/2},~E_{1+}^{3/2}$
 in Eq. 15 at the resonance position;

 (2) the coupling constants $g^V_\omega,~g^T_\omega,~g^V_\rho,~g^T_\rho$
 which describe the $\omega$ and $\rho$ contributions in Eq.~14.

 (3) we have also included into the fitting procedure
the coupling constants for the resonances $S_{11}(1535)$,
 $P_{11}(1430)$ and $D_{13}(1520)$, namely,
 the coupling constants for the multipoles
 $E_{0+}^{1/2},~E_{0+}^{(0)}$ of $S_{11}(1535)$,
for the multipoles $M_{1-}^{1/2},~M_{1-}^{(0)}$
 of $P_{11}(1430)$, and
for the multipoles $E_{2-}^{1/2},~E_{2-}^{(0)},~M_{2-}^{1/2},~M_{2-}^{(0)}$
of $D_{13}(1520)$.
A small variation of these coupling constants around the values
obtained in the analysis of Ref. \cite{VPI97} was allowed.

 The obtained results are presented in Tables 6-8.

\section{The Effective Lagrangian Analysis of the Benchmark Dataset}
\centerline{[ R. M. Davidson ]}
\vskip .5cm

\subsection{The Model}

Details of the effective Lagrangian approach(ELA) to pion photoproduction
may be found in Ref.\cite{dmw}. Here I briefly summarize the main features
of the model. The effective Lagrangian consists of the pseudovector (PV)
nucleon Born terms, $t$-channel $\omega$ and $\rho$ exchange, and
$s$- and $u$-channel $\Delta$(1232) exchanges. At the tree-level, the
amplitude is gauge invariant, Lorentz invariant, crossing symmetric,
and satisfies the LET's for these reactions to order $m_{\pi}/M$. However,
the tree-level amplitude violates unitarity. To unitarize this amplitude,
the tree-level multipoles, ${\cal M}_l^T$, are projected out and unitarized
via a K-matrix approach;
\begin{equation}
{\cal M}_l = {\cal M}_l^T \cos \delta_l {\rm e}^{i\delta_l} \; ,
\end{equation}
where $\delta_l$ is the appropriate $\pi N$ elastic scattering phase shifts.
In practice, this is done only for the s- and p-wave multipoles. In order
to keep multipoles of all $l$ values, the unitarized multipoles are
added to the tree-level CGLN ${\cal F}$'s\cite{cgln} and the tree-level
multipoles are subtracted. To give a simple example, if only $E_{0+}$ is
unitarized, then ${\cal F}_1$ in this model is
\begin{equation}
{\cal F}_1 = {\cal F}_1^T + E_{0+}^T (\cos \delta_0 {\rm e}^{i\delta_0}
-1) \; ,
\end{equation}
where ${\cal F}_1^T$ is the tree-level approximation to ${\cal F}_1$.

After unitarization, one is now ready to fix the parameters of the model.
The $\Delta$ mass, $M_{\Delta}$, and width $\Gamma$ are determined by a fit
to the $P_{33}$ phase shift and are not varied in the photoproduction fit.
The pion and nucleon masses are fixed at 139.6 and 938.9 MeV, respectively.
Although one could look at the sensitivity of the photoproduction data
to the pion-nucleon coupling constant, I keep the PV coupling constant, $f$,
fixed at a value of 1.0. The $\rho$ and $\omega$ are taken to be degenerate
with a mass of 770 MeV, and the $V\pi\gamma$ ($V=\rho$ or $\omega$)
coupling constants are taken from the known radiative decays
$V \rightarrow \pi\gamma$. The Dirac and Pauli-like $VNN$ coupling constants
were allowed to vary in the fit. The remaining parameters of the model
are related to the $\Delta$ interactions. Of most interest are the
two $\gamma N\Delta$ coupling constants, $g_1$ and $g_2$, which are related
to M1 and E2 by
\begin{eqnarray}
M1 &=& {e \over 12M} \left( { k_{\Delta} \over M_{\Delta}M} \right) ^{1/2}
\left\{ g_1  (3M_{\Delta}+M)
 -g_2 {M_{\Delta} \over 2M} (M_{\Delta} -M) \right\}
\; , \nonumber \\
E2 &=& - {e \over 6M} { k_{\Delta} \over (M_{\Delta}+M)}
\left( { k_{\Delta}M_{\Delta} \over M} \right) ^{1/2}
\left\{ g_1
-g_2 {M_{\Delta} \over 2M}  \right\}  \; , \nonumber
\end{eqnarray}
where $k_{\Delta}=(M_{\Delta}^2 -M^2)/(2M_{\Delta})$. The remaining parameters
are the off-shell parameters, X,Y,Z, associated with the $\Delta$ transition
vertices. For example, the $\pi N\Delta$ Lagrangian is of the form
\begin{equation}
{\cal L} \sim \bar{\Delta}^{\mu} ( g_{\mu\nu}+a\gamma_{\mu}\gamma_{\nu})
N \partial^{\nu} \pi + {\rm h.c.} \; ,
\end{equation}
where $a$ depends linearly on Z. Thus, the off-shell parameters essentially
control the relative strength of the $\gamma_{\mu}\gamma_{\nu}$ term
compared to the $g_{\mu\nu}$ term. It should be noted that when calculating
matrix elements involving the off-shell parameters the pole in the
$\Delta$ propagator is canceled. Thus, the contributions involving the
off-shell parameters appear as contact terms. As the off-shell parameters
are fitted to the data, these contact terms can partially compensate
for the lack of strong form factors which might arise, for example, from
pionic dressing of the vertices.

\subsection{Results and Discussion}

The nine parameters of the model were fitted to the low-energy benchmark
dataset consisting of 1287 data points. The total $\chi^2$ was 5203
giving a $\chi^2 /df$ of 4.07. The breakdown of the $\chi^2$ according to
observable is given in Table 9. It is seen that the highest $\chi^2 /n$
are for $d\sigma /d\Omega (\pi^0)$ and T$(\pi^0)$, which was true for the
other analyses also. The best agreement was with the photon asymmetry,
$\Sigma$, for both $\pi^0$ and $\pi^+$.

As the $\chi^2$ in this analysis is slightly larger than in the other
analyses, it is useful to look at the $\chi^2$ breakdown in energy bins.
This is shown in Table 10, where $[x,y)$ and $[x,y]$ have their usual
mathematical meaning. It is seen that the fit to the lowest energy bin
is extremely poor. Most of the data in this energy interval are $\pi^0$
differential cross section data, and a comparison of the fit with these
data shows that the fit has too large of a forward-backward asymmetry as
compared to the data. At these energies, this asymmetry is determined by the
interference between the $E_{0+}$ and the p-wave multipoles. A comparison
of this multipole obtained in this fit with the same multipole obtained
from fits that reproduce these low-energy data shows a significant difference.

\begin{table}[h]
\caption{$\chi^2$ for each observable fitted.}
\begin{center}
\begin{tabular}{|c|r|r|c|}
\hline
OBS & n & $\chi^2$  & $\chi^2 /n$  \\
\hline
$d\sigma /d\Omega (\pi^+)$ & 317 & 988.4 & 3.1 \\
$d\sigma /d\Omega (\pi^0)$ & 354 & 2804.0 & 7.9 \\
$\Sigma (\pi^+)$  & 245 & 388.0 & 1.6 \\
$\Sigma (\pi^0)$ & 192 & 258.1 & 1.3 \\
T$(\pi^+)$  & 107 & 274.9 & 2.6 \\
T$(\pi^0)$  & 72 & 489.6 & 6.8\\
\hline
\end{tabular}
\end{center}
\end{table}

\begin{table}[h]
\caption{$\chi^2$ as a function of energy bin fitted.}
\begin{center}
\begin{tabular}{|c|r|r|r|}
\hline
Interval & n & $\chi^2$  & $\chi^2 /n$  \\
\hline
$[180,210)$ & 64 & 747.7 & 11.7 \\
$[210,240)$ & 137 & 767.6 & 5.6 \\
$[240,270)$  & 208 & 803.6 & 3.9 \\
$[270,300)$ & 181 & 578.3 & 3.2 \\
$[300,330)$  & 191 & 516.8 & 2.7 \\
$[330,390)$  & 157 & 598.6 & 3.8 \\
$[360,390)$  & 120 & 265.5 & 2.2 \\
$[390,420)$  & 132 & 307.6 & 2.3 \\
$[420,450)$  & 97 & 615.4 & 6.3\\
\hline
\end{tabular}
\end{center}
\end{table}

At the BRAG workshop, we were able to understand this difference. As L. Tiator
pointed out, at low-energies for $\pi^0$ production, the coupling to the
$\pi^+ n$ channel can be important and one must somehow account for
processes like
\begin{equation}
\gamma p \rightarrow n\pi^+ \rightarrow p\pi^0 \; .
\end{equation}
In dispersion relation models\cite{ha98,az} and dynamical models\cite{KY99},
this channel coupling is dynamically taken into account. In partial
wave analyses\cite{VPI97}, the parametrization is flexible enough to account
for this physics. In the isobar model\cite{MAID98}, this important physics
was included in a semi-phenomenological manner. In the ELA, this channel
coupling is only partially taken into account via the unitarization. However,
dispersive corrections are not explicitly accounted for. It is implicitly
assumed that the main effect of the dispersive corrections is to renormalize
the parameters to their physical values. It is further assumed (or hoped)
that any additional dispersive corrections can be accounted for by the
contact terms coming from the off-shell parameters. Evidently, this is not
the case over the entire fitted energy range. Thus, to improve the
low-energy fit, without destroying the fit near the peak of the resonance,
this additional physics would need to be added by hand to the ELA.

The parameters from the fit are shown in Table 11 along with representative
parameters from Ref.\cite{dmw}. The resulting resonance couplings are
\begin{eqnarray}
M1 &=& 286.2 \times 10^{-3} {\rm GeV}^{-1/2} \nonumber \\
E2 &=& -7.21 \times 10^{-3} {\rm GeV}^{-1/2} \nonumber \\
{ E2 \over M1} &=& - 2.55 \% \; .
\end{eqnarray}

\begin{table}[h]
\caption{Parameters obtained in this fit (B.M.) compared with those
obtained in \protect\cite{dmw}.}
\begin{center}
\begin{tabular}{|l|r|r|}
\hline
Par. & B.M. & DMW \\
\hline
 $g_1$ & 5.06 & 5.01 $\pm$ 0.22 \\
 $g_2$ & 4.72 & 5.71 $\pm$ 0.43 \\
 Z & -0.25 & -0.30 $\pm$ 0.12 \\
 Y & 1.65 & -0.38 $\pm$ 0.66 \\
 X & -5.74 & 1.94 $\pm$ 2.28 \\
 $g_{\omega 1}$ & -0.09 & 8 \\
 $g_{\omega 2}$ & 4.46 & -8 \\
 $g_{\rho 1}$ & 1.51 & 2.66 \\
$g_{\rho 2}$ & 3.66 & 16.2\\
\hline
\end{tabular}
\end{center}
\end{table}

The $g_1$ coupling, which is mostly responsible for the $M1$ strength,
has not changed much compared to the earlier work. $g_2$, which determines
the strength of $E2$ has changed within the error bar. As a rule of thumb,
the smaller $g_2$, the larger in magnitude $E2$. The change in $g_2$, and
hence $E2$, is probably due to the new high-precision $\Sigma$ data
\cite{legs,beck}. The off-shell parameter Z has not changed much, but the
other parameters have. This is partly due to the fact that the vector meson
couplings were allowed to vary in this fit, but not in Ref.\cite{dmw}.
Since I did not do an error analysis for the BM fit, it is hard to say
if the changes are significant. I do know that the vector meson couplings
and off-shell parameters are highly correlated fit parameters. I should
point out that the vector mesons are put in as point particles, i.e.,
no form factors. Thus, when I fit the vector meson coupling `constants',
I would expect them to decrease compared to their on-shell values. However,
the results for the $\omega$ seem unusual. The general conclusion at the
BRAG workshop is that the $\Delta$ region is not a good place to fit the
vector meson couplings.

From a glance through the various multipole solutions, it seems to me
that the model dependence is under good control in the $\Delta$ region.
One multipole that remains to be understood is the $M_{1-}^{1/2}$. Though
all analyses pretty much agree on its numerical value, its physical
interpretation is quite different in the various models. Is it a
crossed-$\Delta$ effect or the tail of the Roper? In principle, the
amplitude should be crossing symmetric, so if there is an s-channel
$\Delta$ exchange, there must be a u-channel $\Delta$ exchange. In
Ref.\cite{dmw}, it was found that the $M_{1-}^{1/2}$ multipole is largely
insensitive to the off-shell parameters. Thus, these contributions cannot
suppress the u-channel contribution in this multipole, and there is no
room for a Roper contribution as large as in the isobar model. What is
even more mysterious is that in Ref.\cite{dmw} it was found that the
Roper contribution enters with the opposite sign than the u-channel
contribution (see Fig. 8 in \cite{dmw}). Regarding the Roper
contribution to this multipole, there are two clear possibilities.
Either someone has made a mistake, or two different things are being
compared. Certainly the latter is true to some extent. In Ref.\cite{dmw},
both the s- and the u-channel Roper exchanges were included, whereas
in the isobar model\cite{MAID98} only the s-channel contribution was
included. Thus, it is possible that the s- and u-channel contributions
destructively interfere in this multipole to produce the small effect
found in \cite{dmw}. I do not know the solution to this puzzle, but if
dispersion relations are telling us there is a large crossing contribution
to this multipole from the $M_{1+}^{3/2}$, then we must necessarily
investigate this problem within the framework of a crossing symmetric
model.

\section{Multipole Analysis of the Benchmark Set Covering the
First Resonance Region}
\centerline{[ A.S. Omelaenko ]}
\vskip .5cm

\subsection{Introduction}

A great deal of experimental data on the photoproduction of single
pions is stored in compilations. There are problems connected with
the normalization of some systematic measurements of the differential
cross section. As a result, in multipole analyses, one has to reject
or renormalize up to 10\% of experimental points. This can throw
some doubt on the extraction of delicate values, such as the E2/M1
ratio for the $\Delta^+(1232)$.
This makes very interesting the comparison of analyses, based on
different approaches, using the same test database for the
most well investigated $\gamma p\to p\pi^0(n\pi^+)$ reaction, with
a strong accent on the modern data.
Here we report on a fit to the low-energy dataset based
on (a) the resonance model with polynomial parametrization of the
background and (b) energy-independent fitting.

\subsection{Formalism in the First Resonance Region}

{\bf Resonance Model.}
Description of the $P_{33}$ $\pi N$ scattering amplitude is simple
if it is taken to be purely elastic in the first resonance region.
Similar to the form used in Ref.\cite{Walker}, our multi-parameter
model for single pion photoproduction is written as a sum of a
background plus the $\Delta^+ (1232)$ resonance contribution. The
real part of the background is given by the electric Born
approximation which is completed in s-, p- and d-waves by a
cubic polynomial of the form

\begin{equation}
{\rm Re} M_{l\pm}^I (E_{\gamma}) = \sum_{i=1}^4 {\rm Re} M_{l\pm}^I
(E_{\gamma}^{(i)}) \stackrel{j\ne i}{\prod_{j=1}^4} (E_{\gamma}
- E_{\gamma}^{(j)} ) / (E_{\gamma}^{(i)} - E_{\gamma}^{(j)} ).
\end{equation}
Here, depending on the laboratory photon energy $E_{\gamma}$,
multipoles $M_{l\pm}^I \equiv A_{l\pm}^I, B_{l\pm}^I$ are defined
according to Ref.\cite{Walker} with the isospin structure
\begin{eqnarray}
A^{1/2} & = & 1/3 A_{\pi^0 p} + \sqrt{2}/3 A_{\pi^+ n} ,\nonumber \\
A^{3/2} & = & A_{\pi^0 p} - \sqrt{1/2} A_{\pi^+ n} . \nonumber
\end{eqnarray}
Index $l$ is the orbital angular momentum, while $+$ and $-$ correspond to
the total angular momentum $j=l\pm 1/2$. Imaginary parts of the background
multipoles are calculated according to Watson's theorem, using phase
shifts from $\pi N$ elastic scattering analyses:
\begin{equation}
{\rm Im} M_{l\pm}^I = {\rm Re} M_{l\pm}^I \tan (\delta_{2I,2(l\pm)} ).
\end{equation}
The resonant multipole amplitudes $A_{1+}^{3/2}$ and $B_{1+}^{3/2}$ have
the following form
\begin{equation}
M_{1+}^{3/2} (E_{\gamma}) = M_{1+}^{3/2,R}(E_{\gamma}) +
B_M(E_{\gamma}) \cos \delta_{33} e^{i\delta_{33} }
\end{equation}
with the first term describing excitation of the resonance
\begin{equation}
M_{1+}^{3/2,R}(E_{\gamma}) = C_M \sqrt{ {k_0 q_0}\over {k q} }
{{W_0 \sqrt{\Gamma \Gamma_{\gamma}} }\over {W_0^2 - W^2 -i W_0 \Gamma }},
\end{equation}
with $W$ and $W_0$ being the total c.m. energy and its value at resonance,
respectively, $C_M$ being the resonance constant. The energy-dependent widths
were parametrized by
\begin{eqnarray}
\Gamma_{\gamma} (W) & = & \Gamma_0 {\left( {k\over k_0} \right)}^2
\left( { {k_0^2 - X^2}\over {k^2 - X^2} }  \right) ^2, \\
\Gamma (W) & = & \Gamma_0 \left( {q\over {q_0}} \right) ^2
\left( { {q_0^2 - X^2}\over {q^2 - X^2} } \right) ^2
\end{eqnarray}
where $k$ and $q$ are the c.m. momenta of the photon and pion, respectively, and
$k_0$, $q_0$ are the corresponding values at $W_0$. The second term in Eq. (23)
corresponds to Noelle's unitary treatment of the background\cite{Noelle}. In our
approach, the resonance term has a simplified form, not containing the elastic
background phase shift $\delta_B$. However, this has practically no influence
on the E2/M1 ratio, being quite independent of $\delta_B$\cite{omel}. The real
background functions $B_A(E_{\gamma})$ and $B_B(E_{\gamma})$ are parametrized
according to Eq. (21), each in terms of 4 knot values. The phase shift
$\delta_{33}$ is calculated as the phase of the Breit-Wigner form in Eq. (24):
\begin{equation}
\tan \delta_{33} = {{W_0 \Gamma (W) }\over {W_0^2 - W^2}}
\end{equation}
The resonance quantities $C_A$, $C_B$, $W_0$, $\Gamma_0$ were determined in the
fit, along with the knot values of $B_A$ and $B_B$ and the real parts of the
background multipoles: $A_{0+}$, $A_{1+}$, $A_{1-}$, $A_{2-}$, $B_{1+}$, $B_{2-}$
with $I=1/2$ and $A_{0+}$, $A_{1-}$ with $I=3/2$. Inclusion of the d-wave
multipoles is motivated by a possible influence of the second resonance region.

{\bf Energy-Independent Version.}
In order to fit narrow energy bins, we take the real parts of the
abovementioned non-resonant multipoles as independent parameters.
Eq.~(22) is used to calculate imaginary parts, taking into account the
energy dependence of the phase shifts. To avoid calculational problems at
the point where $\delta_{33}$ passes through $\pi / 2$, the resonant multipoles
have been parametrized as
\begin{eqnarray}
A_{1+}^{(3/2)} & = & A \exp (i\delta_{33} ) , \\
B_{1+}^{(3/2)} & = & B \exp (i\delta_{33} ) ,
\end{eqnarray}
with $A$ and $B$ to be determined for each energy bin along with the real
parts of all background multipoles.

\subsection{Results from the Fits}

{\bf Resonance Model.}
The whole energy interval was
split into 3 subintervals, with knot values corresponding to the central
values for energy-independent fits.
\begin{table}[htbp]
\begin{center}
\begin{tabular}{|c|c|c|}
\hline
Fit 1 & 180$\le E_{\gamma} \le$350 MeV & $E_{\gamma}^{(1,2,3,4)}$=200,250,300,350 MeV \\
Fit 2 & 250$\le E_{\gamma} \le$400 MeV & $E_{\gamma}^{(1,2,3,4)}$=250,300,350,400 MeV \\
Fit 3 & 300$\le E_{\gamma} \le$450 MeV & $E_{\gamma}^{(1,2,3,4)}$=300,350,400,450 MeV \\
\hline
\end{tabular}
\end{center}
\end{table}

\begin{table}[htbp]
\begin{center}
\begin{tabular}{|c|c|c|c|c|}
\hline
Fit &  Data & $\chi^2$ & N & $\chi^2$/N \\
\hline
Fit 1 &  Total & 2026.5 &  930  & 2.18 \\
\hline
              &   $\gamma p\to p\pi^0$ & 1332.4 & 460  & 2.90 \\
\hline
              &   $\gamma p\to n\pi^+$ & 694.1  & 470  & 1.48 \\
\hline
Fit 2  & Total & 1759.2 & 871 & 2.02 \\
\hline
              &  $\gamma p\to p\pi^0$ & 1115.7 & 388 & 2.88 \\
\hline
              &  $\gamma p\to n\pi^+$ & 643.5  & 483 & 1.33 \\
\hline
Fit 3  & Total & 1259.7 & 697 & 1.81 \\
\hline
              &  $\gamma p\to p\pi^0$ & 821.6 & 306 & 2.69 \\
\hline
              &  $\gamma p\to n\pi^+$ & 438.1 & 391 & 1.12 \\
\hline
\end{tabular}
\end{center}
\caption{Fitting in the resonance model framework.}
\end{table}

\begin{table}[htbp]
\begin{center}
\begin{tabular}{|c|c|c|c|c|}
\hline
Reaction & Observable & Label & N & $\chi^2$/N \\
\hline
$\gamma p\to p\pi^0$ & $d\sigma /d\Omega$ & FU96MA & 45 & 2.2 \\
\hline
       & $d\sigma /d\Omega$ & HA97MA & 52 & 10.5 \\
       & $d\sigma /d\Omega$ & BE97MA & 77 & 1.4 \\
       & Sigma  & BL92LE & 16 & 3.3 \\
       & Sigma  & BE97MA & 77 & 0.8 \\
       & T      & BO98BO & 28 & 3.1 \\
       & T      & BE83KH & 26 & 3.6 \\
\hline
$\gamma p\to n\pi^+$ & $d\sigma /d\Omega$ & FI72BO & 32 & 0.9 \\
\hline
       & $d\sigma /d\Omega$ & BR00MA & 39 & 1.9 \\
       & $d\sigma /d\Omega$ & BE00MA & 140 & 1.7 \\
       & Sigma & BL00LE & 57 & 1.7 \\
       & Sigma & BE00MA & 140 & 0.9 \\
       & T & DU96BO & 75 & 1.1 \\
\hline
\end{tabular}
\end{center}
\caption{Fits with the resonance model in the 250-400 MeV interval. See the
database summary webpage to match abbreviations with full references.}
\end{table}

\begin{table}[htbp]
\begin{center}
\begin{tabular}{|c|c|c|c|}
\hline
Interval (MeV) & Mass (MeV) & Width (MeV) & E2/M1 (\%) \\
\hline
180-350  & 1232.6 $\pm$ 1.0 & 121.1 $\pm$ 6.0 & -1.83$\pm$0.31 \\
\hline
250-400  & 1231.7$\pm$ 0.7 & 118.4 $\pm$ 3.9 & -2.34$\pm$0.13 \\
\hline
300-450 & 1230.6 $\pm$ 0.8 & 116.5 $\pm$ 3.9 & -2.26$\pm$0.14 \\
\hline
\end{tabular}
\end{center}
\caption{Resonance parameters for the $\Delta (1232)$ from different energy
intervals.}
\end{table}
For each fit a total of 44 free parameters were searched (the parameter $X$ turned out
to be difficult to determine, and was fixed at Walker's value of 185 MeV). The statistical
characteristics for all fits are given in Table 12, separately for the
neutral and charged particle production. In Table 13, the contribution to $\chi^2$ from
individual experiments is given, as calculated from Fit 2. Finally, in Table 14, values
for the resonance mass and width are given along with the corresponding value for the
E2/M1 ratio.

{\bf Energy-Independent Analysis.}
Results from the resonance model fits were used as starting values for 10 parameters,
which were determined by $\chi^2$ minimization from the data contained in
each 10 MeV bin. For all formally successful fits, the corresponding central values
for photon energy, number of points (N) and $\chi^2$ per degree of freedom is given
in Table~15.

\begin{table}[htbp]
\label{tab:enind}
\begin{center}
\begin{tabular}{|c|c|c|c|c|c|c|c|}
\hline
$E_{\gamma}$, MeV  &$\chi^2$ &N & $\chi^2/df$ &
$E_{\gamma}$, MeV  &$\chi^2$ &N & $\chi^2/df$  \\
\hline
210   &10.1  &18    &1.27  &330   &100.6 &51    &2.45\\
220   &18.6  &43    &0.56  &340   &20.0  &49    &0.51\\
230   &23.8  &22    &1.99  &350   &118.3 &74    &1.85\\
240   &52.9  &58    &1.10  &360   &6.3   &32    &0.29\\
260   &112.3 &71    &1.84  &370   &28.9  &42    &0.90\\
270   &33.3  &36    &1.28  &380   &93.1  &46    &2.59\\
280   &171.5 &69    &2.91  &390   &16.7  &42    &0.52\\
290   &72.9  &51    &1.78  &400   &86.8  &56    &1.89\\
300   &81.3  &80    &1.16  &410   &16.7  &34    &0.70\\
310   &121.8 &59    &2.49  &420   &67.2  &50    &1.68\\
320   &96.8  &69    &1.64  &      &      &      & \\
       \hline\hline
\end{tabular}
\caption{Statistics of the energy-independent fits.}
\end{center}
\end{table}

\subsection{Concluding Remarks}

In the low-energy region, dominated by the first resonance, we find
a rather good determination of the main s- and p-wave partial-wave
amplitudes for pion photoproduction (on proton targets), taking
into account d-wave corrections. Differences between the Fits 1-3
on overlapping energy intervals can be considered as some measure
of the model error. In our analysis, a suspicious deviation in
the energy dependence of the $M_{1-}^{1/2}$ and $M_{1+}^{1/2}$
was evident over the 380-420 MeV interval. Encouraging was the
rather stable determination of the mass and width of the
$\Delta^+ (1232)$ resonance and E2/M1 ratio within the framework
of this model (Fits 2 and 3).

A comment of the chosen database is in order. Of the cross
section data for neutral pion production, the set of HA97MA
gives a $\chi^2$ which is definitely too large. This is
true in all of the fits. Problems seem concentrated mainly
at the small and large angles. Unfortunately, the modern
polarization data do not represent the complete set of
single polarization observables. As a result, in some bins
of the energy-independent fit, the experimental data were
not sufficient for a successful minimization procedure,
or yielded unreasonably low (less than 1) values for
$\chi^2/df$.
Further examination has shown that the abovementioned
deviations in the energy behavior of the d-wave multipoles
are due to the lack of reliable data as well.
Precise new experiments would be extremely desirable.

\section{SAID Multipole Analysis of the Benchmark Dataset}

\centerline{[ R.A. Arndt, I.I. Strakovsky, R.L. Workman ]}
\vskip .5cm

SAID fits were performed on both the low- and medium-energy datasets,
using a phenomenological method described in Ref.\cite{VPI97}. Multipoles
were parametrized in the form
\begin{equation}
M = \left( A_{\rm phen} + Born \right) (1 +i T_{\pi N} )
  +        B_{\rm phen} T_{\pi N}
\end{equation}
where $A_{\rm phen}$ and $B_{\rm phen}$ were constructed from polynomials
in energy having the correct threshold behavior. An additional overall
phase proportional to ${\rm Im} T_{\pi N} - T_{\pi N}^2$ was also allowed in
order to allow deviations from the above form for waves strongly
coupled to channels other than $\pi N$. This form clearly satisfies
Watson's theorem, where valid, and allows a smooth departure as
new channels become important. A real Born contribution is assumed for
unsearched high partial-waves.

The overall fit to data is summarized in the Table below. While the
fit to most quantities is about 2 per data point, there are clear
exceptions. Our fit to the neutral-pion differential cross section
and target polarization is significantly worse in the low-energy
region. If $\chi^2$ is calculated for the region between 450-1200 MeV
a fairly uniform fit emerges, with all data types fitted equally
well.

\begin{table}[htbp]
\begin{center}
\begin{tabular}{|c|c|c|c|}
\hline
Observable & Low-Energy Fit & Medium-Energy Fit & Medium-Energy Fit \\
           & (180-450)MeV   & (180-1200)MeV   & (450-1200) MeV \\
\hline
$\sigma (\theta)_{\pi^0 p}$ & 4.69   &   3.14   &   1.80  \\
\hline
$\Sigma (\theta)_{\pi^0 p}$ & 1.24   &   2.05   &   2.32  \\
\hline
$T(\theta)_{\pi^0 p}$       & 4.49   &   2.84   &   1.89  \\
\hline
$\sigma (\theta)_{\pi^+ n}$ & 2.12   &   2.55   &   2.08  \\
\hline
$\Sigma (\theta)_{\pi^+ n}$  & 1.46   &   1.81   &   2.36  \\
\hline
$T(\theta)_{\pi^+ n}$       & 2.92   &   2.25   &   1.83  \\
\hline
\end{tabular}
\end{center}
\caption{Comparison of the low- and medium-energy fits
in terms of $\chi^2$.}
\end{table}

The poor fit to neutral-pion differential cross sections can
be traced to the Haerter data from Mainz (Ph.D. Thesis, unpublished -
see the database description on the BRAG webpage). This set contributed
a $\chi^2$/datum of 10, the remaining data again having an overall
$\chi^2$ per data point of about 2.

Why the fit to target polarization should be poor is harder to
determine. In the low-energy region, the benchmark dataset has
taken target-polarization data from Kharkov and Bonn. Of these,
the fit to Kharkov data is particularly bad. Our fit to the Bonn data is
reasonable apart from the lowest-energy (and perhaps also the highest
energy) set.

\section{Conclusions and Future Projects}

\subsection{Database Issues}

As this exercise was motivated by an earlier study of the E2/M1 ratio,
and its database dependence, we should (at least) expect to verify the
relative model-independence of this quantity (when a standard dataset
is selected). This point was discussed  by Davidson\cite{davidson} in
his contribution to NSTAR2001. From the combined set of
benchmark fits, the E2/M1 ratio was found to be $-2.38\pm 0.27\%$.
Here the important quantity is the error, since a change in the fitted
dataset could shift the central value.

Qualitative features of the fit to data were reasonably similar in
the low-energy region. There was general agreement that the differential
cross section and target polarization were badly fitted in the
neutral-pion channel. In all cases, the thesis data of Haerter were
identified as problematic for the differential cross section.
[These data can be abandoned, as a new Mainz measurement\cite{Leukel}
of the cross section and photon asymmetry for neutral-pion photoproduction
has been analyzed and is nearing publication. This new set covers the
full angular range and a wide range of energies.]

Unfortunately,
such a simple solution was not evident for the target polarization.
The fit to both the Bonn and Kharkov data is poor, with some sets
suggesting a different shape. In this case, the general {\it disagreement}
should motivate a re-measurement of this quantity, hopefully with
better precision. A confirmation of the shape suggested by these data
would be very difficult to accommodate.

\subsection{Model Dependence}

One ingredient common to all analyses is the Born contribution, usually
augmented with vector-meson exchange diagrams. While it is tempting to
determine an optimal set of vector meson couplings,
over the fitted energy range,
our study has shown this to be a very unstable procedure,
when fits are restricted to the low-energy database.

The $M_{1-}^{1/2}$ multipole, and its low-energy behavior, provides an
interesting example of the model dependence encountered when one
tries to decompose a partial-wave into its underlying ingredients.
Hanstein has noted that, in his dispersion integral, the coupling
$M_{1+}^{3/2} \to M_{1-}^{1/2}$ exceeds
$M_{1-}^{1/2} \to M_{1-}^{1/2}$. In the study of Davidson, this
is accounted for by a u-channel $\Delta$ exchange, a contribution
absent in many isobar-model approaches.
As a result, the Roper resonance contribution depends strongly on
the set of approximations defining a fit.

\subsection{Further Comparisons}

One unique feature of this study was the construction of a site\cite{BRAG}
containing, in one place, both the database and the multipole fits.
This facility was built into the SAID site\cite{VPI97}, thus allowing
a wide range of comparisons. Users can compare fits to observables
(in both the benchmark and full database), the resulting multipoles,
and search for regions of maximal agreement or disagreement.

\subsection{Extensions of this Study}

Having the present set of fits as a guide, it would be useful to
repeat this study over a more carefully chosen database - containing,
in some way, the effects of systematic errors. A second project of
interest would be the extension to electroproduction, with the goal
of understanding the model dependence seen in the extracted
E$_{1+}$/M$_{1+}$ and S$_{1+}$/M$_{1+}$ ratios as a function of $Q^2$.

\end{document}